\documentstyle[aps,prl]{revtex}

\begin{document}

\twocolumn[\hsize\textwidth\columnwidth\hsize\csname
@twocolumnfalse\endcsname

\draft{}
\title{Spin Torques in Ferromagnetic/Normal Metal Structures}
\author{K. Xia$^1$, P. J. Kelly$^1$, G. E. W. Bauer$^2$, A. Brataas$^3$, and I. Turek$^4$}
\address{$^{1}$Faculty of Applied Physics and MESA$^{+}$ Research Institute,\\
University of Twente, P.O. Box 217,\\
7500 AE Enschede, The Netherlands\\
$^2$Department of Applied Physics and DIMES, Delft University of Technology,\\
Lorentzweg 1, 2628 CJ Delft, The Netherlands\\
$^3$Harvard University, Lyman Laboratory of Physics, Cambridge, MA 02138\\
$^{4}$Institute of Physics of Materials, Academy of Sciences of the Czech\\
Republic, CZ-616 62 Brno, Czech Republic}
\date{\today}
\maketitle

\begin{abstract}
Recent theories of spin-current-induced magnetization reversal are
formulated in terms of a spin-mixing conductance $G^{mix}$. We evaluate $G^{mix}$ from first-principles for a number of (dis)ordered interfaces
between magnetic and non-magnetic materials. In multi-terminal devices, the
magnetization direction of a one side of a tunnel junction or a
ferromagnetic insulator can ideally be switched with negligible charge
current dissipation.
\end{abstract}


\vskip2pc]

\narrowtext

``Giant magnetoresistance'' refers to the large change of resistance brought
about by applying an external magnetic field so as to change the angle
between the magnetization directions of adjacent magnetic films\cite
{GMR_reviews}. Since a spin injected into a magnetic material experiences a
torque, it has been argued that there should be a reverse effect, namely,
that passage of a current through adjacent magnetic layers should lead to
the transfer of spin angular momentum from one layer to the other\cite
{Slon96,Berger96} with possible reorientation of the magnetizations for
sufficiently large currents \cite{Slon96}. Interestingly, the sign of the
corresponding torque should be reversed on changing the current direction
leading to the possibility of making an electronically accessible
non-volatile magnetic memory whose performance on downscaling would appear
to compare favorably with other alternatives\cite{Inomata01}. The
experimental observation of current-induced magnetization reversal (``spin
transfer'') effects by a number of groups\cite
{Tsoi98,Wegrowe99,Sun99,Myers99,Katine00} calls for a more general framework
for electrical transport in non-collinear and disordered magnetic systems.
Two suitable and nearly equivalent approaches\ are the circuit theory based
on kinetic equations for spin currents \cite{Brataas00} and the random
matrix theory of transport in magnetic heterostructures \cite{Waintal00}.

In the circuit theory, the torque and current are formulated in terms of
(real) spin-up and spin-down conductances $G^{\uparrow }$ and $G^{\downarrow
}$ and a new spin-mixing conductance $G^{mix}$ which is complex. The mixing
conductance has only been studied using simple free electron models and
little is known about its dependence on real material parameters. Since
free-electron models miss an important contribution to spin transport in
layered magnetic materials coming from the mismatch of the complex $d$ bands
responsible for itinerant ferromagnetism\cite{Schep95,Zahn95}, it is
important to take this into account when evaluating $G^{mix}$ for realistic
interfaces.

In this paper we use methods recently developed \cite{Schep97,Stiles00,Xia01}
to calculate the scattering matrix within the framework of density
functional theory in order to evaluate the spin- and spin-mixing
conductances for a number of systems of current interest. We show that the
spin-mixing conductance in tunnel junctions can remain large even when the
conventional conductance itself is made vanishingly small, which has not
been anticipated previously and should be important in utilizing the
spin-torque effects. We suggest a configuration in which this effect could
be observed.

We begin with two basic elements of a magnetic circuit: a non-magnetic (NM)
element in which there is a spin-accumulation in the direction ${\bf s}$ [as
a result of injection from elements of the circuit e.g. by a source-drain
current in Fig.~\ref{Fig1}(a)] and a ferromagnetic (FM) element whose
magnetization is given by the unit vector ${\bf m}$. In general ${\bf s}$
and ${\bf m}$ are non-collinear and it is convenient to split the $2\times 2$
matrix current $\hat{I}$ from the non-magnetic element into the
ferromagnetic element into a scalar charge current $I_{0}$ and a vector spin
current ${\bf I}_{s}$: $\hat{I}=(I_{0}+\text{\boldmath$\sigma $}\cdot {\bf I}%
_{s})/2$ where $\text{\boldmath$\sigma $}$ is a vector of Pauli spin
matrices, 
\begin{eqnarray}
{\bf I}_{s} &=&{\bf m}[(G^{\uparrow }-G^{\downarrow
})(f_{p}^{F}-f_{p}^{N})+(G^{\uparrow }+G^{\downarrow })\Delta f^{F} 
\nonumber \\
&&+(G^{\uparrow }+G^{\downarrow }-2\text{Re}G^{mix}){\bf s}\cdot {\bf m}%
\Delta f^{N}]  \nonumber \\
&&-2{\bf s}\text{Re}G^{mix}\Delta f^{N}+2({\bf s}\times {\bf m})\text{Im}%
G^{mix}\Delta f^{N}\,,  \label{vsc}
\end{eqnarray}
and $\hat{f}^{N}=f_{p}^{N}+\text{\boldmath$\sigma $}\cdot {\bf s}\Delta
f^{N} $ and $\hat{f}^{F}=f_{p}^{F}+\text{\boldmath$\sigma $}\cdot {\bf m}%
\Delta f^{F}$ describe the deviation of the distribution functions in the NM
and FM nodes, respectively, from their equilibrium values\cite{NEC01}. When
a spin-current is injected into a ferromagnetic material, the component of $%
{\bf I}_{s}$ perpendicular to the magnetization direction ${\bf m}$ (times
the Bohr magneton and gyromagnetic ratio) equals the torque acting on the
ferromagnet \cite{Waintal00} and this is seen to be determined entirely by
the real and imaginary parts of $G^{mix}$. The spin-up, spin-down and
spin-mixing conductances characterizing transport through an interface are
defined as $G^{\uparrow }=\frac{e^{2}}{h}tr(t_{\uparrow }^{\dagger
}t_{\uparrow })$, $G^{\downarrow }=\frac{e^{2}}{h}tr(t_{\downarrow
}^{\dagger }t_{\downarrow })$ and $G^{mix}=\frac{e^{2}}{h}tr(I-r_{\uparrow
}^{\dagger }r_{\downarrow })$, in terms of the spin-dependent transmission
and reflection matrices $t_{\uparrow (\downarrow )}$ and $r_{\uparrow
(\downarrow )}$; $I$ is an $M\times M$ unit matrix where $M$ is the number
of conducting channels in the NM element \cite{Brataas00,Waintal00}. Note
that the mixing conductance can be measured in principle by the angular
magnetoresistance of perpendicular spin valves \cite{Pratt} analyzed by
circuit theory \cite{Huertas01a}.

The parameter-free calculation of the transmission and reflection matrices 
\cite{Xia01} is based on the surface Green's function method \cite{Turek}
implemented with a tight-binding linear muffin tin orbital basis\cite
{Andersen85}. The calculation time scales linearly with the number of layers
in the scattering region and as the cube of the size of the in-plane unit
cell. Because a minimal basis set is used, we are able to perform
calculations for lateral supercells containing as many as $10\times 10$
atoms and to model disorder very flexibly within such supercells without
using any adjustable parameters. The electronic structure is determined
self-consistently within the local spin density approximation. For
disordered layers the potentials are determined self-consistently using the
layer CPA approximation \cite{Turek}. The calculations are carried out with
a ${\bf k_{\parallel }}$ mesh density equivalent to more than 3600 mesh
points in the two-dimensional Brillouin zone (BZ) of a $1\times 1$ interface
unit cell. The results of calculations for clean and disordered fcc(111)
Cu/Co and bcc(001) Cr/Fe metallic interfaces and for an fcc(111)
Cu/Co/Vac/Co tunneling configuration are given in Table~\ref{tabone}.

We first discuss our results for the Cu/Co and Cr/Fe interfaces. Both have
been the subject of much study in the context of exchange coupling and giant
magnetoresistance and different calculations of the interface transmission
matrices yield very similar results for the spin-dependent interface
resistances\cite{Schep97,Stiles00,Xia01}. The atomic volume of each pair of
materials is very similar and we model disordered interfaces using a 2 layer
thick 50-50 alloy of the component materials. The results given in the Table
do not depend sensitively 
on the alloy concentration used.

For a clean Cu/Co(111) interface, the real part of the mixing conductance is
comparable in size to the spin-up and spin-down conductances but the
imaginary part is almost a factor of 50 smaller. Interface disorder
increases the mixing conductance, the real part by about 35\%, the imaginary
part by a factor of three.

It is interesting to compare $|G^{mix}|$ with $G^{\uparrow }+G^{\downarrow }$%
. The mixing conductance mainly contributes to the torque while $G^{\uparrow
}+G^{\downarrow }$ determines the electron current. Large values of $%
|G^{mix}|/(G^{\uparrow }+G^{\downarrow })$ mean more torque per unit
current. The calculations show that disorder at the Co/Cu interface
increases the spin-torque. Another possibility to increase the ratio of $%
|G^{mix}|$ to $G^{\uparrow }+G^{\downarrow }$ is to insert an impurity layer
on the Cu side; a Co monolayer inserted on the Cu side scarcely changes the
mixing conductance but reduces the normal conductance significantly.

For the Cr/Fe interface the band structure matching and the effect of
interface disorder are quite different compared to Cu/Co. Whereas the
majority spin states of Co/Cu match very well, it is the minority
spin-states in Cr and Fe which match best; see Figs.~\ref{Fig1}(c,d). For a
perfect Cr/Fe(001) interface, the mixing conductance is almost twice as
large as the normal conductance. The expression for the mixing conductance
shows that having a large number of propagating channels on the
non-ferromagnetic side of the interface can lead to a large mixing
conductance. 

It is of interest to have a closer look at the ${\bf k_{\parallel }}$
resolved mixing conductance$\,G^{mix}$. We can see from Fig.~\ref{Fig1}(e)
that close to the center of the Brillouin zone (BZ) the real part of the $%
G^{mix}$ is very large, even larger than the number of channels in Cr at the
same ${\bf k_{\parallel }}$-points, shown in Fig.~\ref{Fig1}(b); at the same 
${\bf k_{\parallel }}$-points the transmission of majority spin electrons is
very low. Thus at some ${\bf k_{\parallel }}$ the mixing conductance can be
much larger than the normal conductance. This can be understood in terms of
a simple one-dimensional infinite barrier model in which the spin-up and
spin-down barriers are displaced in space by an amount $\Delta $. For both
spins the reflection amplitude is $1$. However, the displacement $\Delta $
introduces a phase shift $e^{-2ik\Delta }$ for electrons with wave-number $k$
so that the mixing conductance can have any value $G_{0}(1-e^{-2ik\Delta })$
for this simple single-channel model. Although $G^{mix}$ is large around $%
{\bf k_{\parallel }}=0$ for Fe/Cr, the minority-spin reflection is very low
in most parts of the BZ so that after averaging over the BZ, $G^{mix}$ is
not very high compared with the normal conductance.

The imaginary part of $G^{mix}$ is related to the spin precession which
results from the non-collinear alignment of the spins of the injected
electrons and the magnetization (or an external magnetic field). A
non-vanishing imaginary part of the mixing conductance, Im$\,G^{mix}$,
should result in antisymmetry with respect to time reversal\cite
{Huertas00,Huertas01b}. However, Im$\,G^{mix}$ is small in all the systems
we have studied. The reason for this can be understood by examining the $%
{\bf k_{\parallel }}$ resolved imaginary part of $G^{mix}$ shown in Fig.~\ref
{Fig1}(f). Im$\,G^{mix}$ can be negative as well as positive and it is the
partial cancellation of these contributions which leads to the net result
being small. This can be illustrated using the simple phase-shift model as
follows. Suppose that the phase shift of the reflected waves $\delta ({\bf %
k_{\parallel }})$ is distributed randomly between $\varphi _{1}$ and $%
\varphi _{2}$ with equal weights and that the amplitude of $r_{\uparrow
}^{\dagger }r_{\downarrow }$ is $A$, then the average mixing conductance is $%
\frac{G_{0}}{\varphi _{2}-\varphi _{1}}\int_{\varphi _{1}}^{\varphi
_{2}}(1-Ae^{i\delta })d\delta =\frac{e^{2}}{h}\left[ 1+\frac{iA}{\varphi
_{2}-\varphi _{1}}(e^{i\varphi _{2}}-e^{i\varphi _{1}})\right] $ where $%
G_{0}=\frac{e^{2}}{h}$. If $\varphi _{1}=-\varphi _{2}$ then$\,%
\mathop{\rm Im}%
G^{mix}$ is zero and $%
\mathop{\rm Re}%
G^{mix}=G_{0}$. In a more realistic treatment the weights will not be
homogeneous and $\varphi _{1}$ will be slightly different from $-\varphi
_{2} $ 
so that $%
\mathop{\rm Im}%
G^{mix}$ is not exactly zero. To obtain a large $%
\mathop{\rm Im}%
G^{mix}$ experimentally, one should use a material with a small number of
conducting channels such as a semiconductor as the non-magnetic element.
Also in a ferromagnetic insulator $%
\mathop{\rm Im}%
G^{mix}$ could be significant \cite{Huertas01b}.

We may conclude that a large reflection amplitude does not mean that the
mixing conductance must be small. It provides us with a means to realize
large values of $|G^{mix}|/(G^{\uparrow }+G^{\downarrow })$ by using a
tunnel junction or ferromagnetic insulator as the FM element. For such
non-conducting interfaces we predict that it will be possible to obtain a
non-zero spin current while the electronic current is zero or vanishingly
small. We confirm this by calculating the mixing conductance for a
Cu/Co/Vac/Cu tunnel junction, here we use 6 layers empty spheres to model
the barrier. The mixing conductance is found to be $0.41-i\,0.04$ and $%
0.53-i\,0.003$ for clean and dirty systems, respectively, in units of $%
10^{15}\Omega ^{-1}m^{-2}$, which very close to the mixing conductances of
Cu/Co interfaces. The normal conductance is of the order of $10^{-11}$ in
the same system. For even thicker layers\ the torque simply equals that of
the metallic interface and the tunnel junction only suppresses the charge
current.

To inject a spin-current in the absence of an electron current we need to
consider a three terminal device (``spin-flip transistor'') such as that
sketched in Fig.~\ref{Fig1}(a) \cite{Brataas00,SFT}. A current from FM1 into
FM2 induces a spin-accumulation in the NM node which is easily calculated 
\cite{SFT}. Here we simply assume that a spin-accumulation exists in NM and
analyze what happens when FM is a magnetic insulator or is the top magnetic
element of a magnetic tunnel junction. In the latter case it is possible to
independently determine the orientation of FM by means of the tunneling
magnetoresistance \cite{Moodera95}. The spin torque is that of the metallic
junction, but without the energy dissipation caused by the particle current.
In practical memory devices it may be advantageous to be able to achieve
this separation of particle and spin injection.

In summary, we have studied the mixing conductance of Cu/Co, Cr/Fe and
Cu/Co/Vacuum/Co configurations taking the full transition metal electronic
structure into account and including disorder. The effect of the mixing and
normal conductances can be separated for a three terminal device where one
of the terminals is a ferromagnetic insulator or a magnetic tunnel junction
for which the normal conductance can be made vanishingly small without
affecting the size of the mixing conductance.

We acknowledge helpful discussions with Yuli Nazarov and Daniel
Huertas-Hernando. This work is part of the research program for the
``Stichting voor Fundamenteel Onderzoek der Materie'' (FOM). This study was
supported by the Norwegian Research Council, and the Grant Agency of the
Czech Republic (202/00/0122). We acknowledge benefits from the TMR Research
Network on ``Interface Magnetism'' under contract No. FMRX-CT96-0089
(DG12-MIHT).

\begin{figure}[h]
\caption{ (a) Sketch of a three-terminal device where a normal metal (NM)
element is connected to three ferromagnetic elements FM1, FM2 and FM. An
applied bias causes a current to flow between FM1 and FM2. If FM is a tunnel
junction or a magnetic insulator then the particle flow into FM should be
vanishingly small. (b) number of propagating channels in first Brillouin
Zone for bulk Cr. For a clean Cr/Fe interface (c) and (d) show the
minority-spin and majority-spin conductances, (e) and (f) the real and
imaginary parts of the mixing conductance. Units of conductance are $e^{2}/h$%
. The result of integrating over the whole Brillouin Zone is given in
brackets at the top of each panel. Note that the values represented by color
differ per panel but with the zeros are represented by the green color.}
\label{Fig1}
\end{figure}

\widetext

\begin{table}[tbp]
\begin{center}
\begin{tabular}{cccccc}
system & interface & $G^{maj}$ & $G^{\min }$ & ${\rm Re}G^{mix}$ & ${\rm Im}%
G^{mix}$ \\ \hline
Cu/Co(111) & Clean & 0.42 & 0.36 & 0.41 & $9.2\times 10^{-3}$ \\ 
Cu/Co(111) & $2\times $ 50-50 alloy & 0.42 & 0.33 & 0.55 & $3.0\times
10^{-2} $ \\ 
Cu/Co(1)/Cu(7)/Co(111) & Clean & 0.40 & 0.22 & 0.41 & $3.2\times 10^{-2}$ \\ 
Cu/Co(1)/Cu(7)/Co(111) & $2\times $ 50-50 alloy & 0.41 & 0.21 & 0.55 & $%
3.6\times 10^{-2}$ \\ 
Cr/Fe(001) & clean & 0.14 & 0.35 & 0.61 & $2.8\times 10^{-2}$ \\ 
Cr/Fe(001) & $2\times $ 50-50 alloy & 0.26 & 0.34 & 0.61 & $5.2\times
10^{-1} $ \\ \hline
Cu/Co/Vac/Co & clean & $9.3\times 10^{-12}$ & $1.9\times 10^{-11}$ & 0.41 & $%
-4.1\times 10^{-2}$ \\ 
Cu/Co/Vac/Cu & $2\times $ 50-50 alloy & $3.3\times 10^{-11}$ & $3.0\times
10^{-11}$ & 0.53 & $3.3\times 10^{-3}$%
\end{tabular}
\end{center}
\caption{Interface conductances in units of $10^{15}\Omega ^{-1}m^{-2}\ $.}
\label{tabone}
\end{table}

\end{document}